\documentclass[12pt,a4paper,]{article}
\usepackage{CJKutf8}
\usepackage{amsmath}
\usepackage{amssymb}
\usepackage[top=2cm, bottom=2cm, left=2cm, right=2cm]{geometry}
\usepackage{indentfirst}
\usepackage{hyperref}
\usepackage[numbers, sort&compress]{natbib}
\usepackage{color}
\newcommand{\boldnabla}{  \nabla \hspace{-0.12in}{\nabla}}

\allowdisplaybreaks[4]

\begin{document}
\begin{CJK}{UTF8}{gkai}

\begin{titlepage}

\begin{flushright}
USTC-ICTS-16-02
\end{flushright}

\vspace{10mm}
\begin{center}
{\Large\bf Covariant Quantization of BFNC Super Yang-Mills Theories and Supergauge Invariance}
\vspace{16mm}

{\large Xu-Dong Wang\footnote{E-mail address: xudwang@ustc.edu.cn}

\vspace{6mm}
{\normalsize \em Interdisciplinary Center for Theoretical Study\\
 University of Science and Technology of China, Hefei, Anhui 230026, People's Republic of China}}

\end{center}

\vspace{10mm}
\centerline{{\bf{Abstract}}}
\vspace{6mm}

To construct renormalizable gauge model in Bosonic-Fermionic noncommutative (BFNC) superspace, we replace the ordinary products of super Yang-Mills model by BFNC star products. To study the renormalization property of the deformed action, we obtain the one-loop 1PI effective action by using background field method at the first order of  BFNC parameters. We also verify the BFNC supergauge invariance of the effective action.  Because there are new terms in effective action, the deformed action is not renormalizable. This imply that additional terms should be added to the deformed action.

\vskip 20pt
\noindent
{\bf PACS Number(s)}: 12.60.Jv, 11.30.Pb, 11.10.Nx, 11.10.Gh


\end{titlepage}

\section{Introduction}
\label{Introduction}

In the recent years there are many studies in noncommutative spacetime~\cite{Douglas:2001ba, Szabo:2001kg}. If we introduce background fields in string theory, the spacetime coordinates will become noncommutative~\cite{Seiberg:1999vs}. Because consistent description of string theory need supersymmetry, it is natural to give noncommutative superspace~\cite{de Boer:2003dn, Ooguri:2003qp, Ooguri:2003tt, Berkovits:2003kj}. There are some works studying field theory in non-anticommutative (NAC) superspace~\cite{Seiberg:2003yz}, such as NAC  Wess-Zumino model~\cite{Britto:2003aj, Grisaru:2003fd, Britto:2003kg, Romagnoni:2003xt} and NAC Super Yang-Mills model~\cite{Grisaru:2005we}. The renormalizability constraints the possible form of these model. There are also some interest in constructing field theory in other kind of noncommutative superspace. In Ref.~\cite{Miao:2013a, Miao:2014mia}, the authors have constructed a renormalizable Wess-Zumino model in Bosonic-Fermionic noncommutative (BFNC) superspace. To construct field theory which have gauge symmetries, in this paper we try to construct super Yang-Mills model in BFNC superspace.

To construct model in noncommutative spacetime, we can start from a model in commutative spacetime  and replace the ordinary products by star products, then study the renormalization property of the deformed model. We can obtain the 1PI effective action of the deformed model by using background field method~\cite{Gates:1983nr} and only keep the divergent part. If there are new terms in the effective action, we could add them to the deformed action and obtain a new action. We continue to study the effective action of the new action until there are not any new terms in the effective action. Because in this method we always do one-loop calculation, the action can only be renormalizable at one-loop level. In order to obtain all order renormalizable action, we can introduce $U(1)$ global symmetries. If we assume the effective action also have these $U(1)$ global symmetries, we can determine all of the divergent operators in the effective action by using dimensional analysis. If we can construct action from these divergent operators, then it will be renormalizable to all order in perturbative theory because it contains all of the divergent terms. It will make the task more easy. This method have been used to construct renormalizable model in NAC and BFNC superspace.

For gauge theories it is not straightforward to construct supergauge invariant action by using the divergent operators. Because the form of gauge transformation was changed after introducing noncommutative star products. To construct renormalizable gauge theory in noncommutative superspace, we can at first calculate the effective action by using background field method. The effective action will be supergauge invariant. By using the effective action we can construct renormalizable gauge theory.

Based on the above consideration, we start from the ordinary super Yang-Mills model in this paper. By introducing BFNC star product we obtain deformed action. Then we calculate the 1PI effective action by using background field method. We can check that the effective action is invariant under noncommutative supergauge transformation. As we expect, because the noncommutative star product, there are new terms in the effective action. This imply that to construct renormalizable gauge model we should add new terms to the deformed action. The organization of this papers is as follows, in section~\ref{BFNC superspace} we define the BFNC superspace and the BNFC star product, in section~\ref{review} we review super Yang-Mills model in NAC superspace and the background field method, in section~\ref{effective action} we obtain the effective action of deformed super Yang-Mills in BFNC superspace and verify its supergauge invariance, in section~\ref{conclusion} we give our conclusion and outlook.

\section{BFNC superspace}
\label{BFNC superspace}

BFNC superspace can be defined as truncation of euclidean ${\cal N}=(1,1)$ superspace. We introduce nonstandard hermitian conjugation rules for the spinorial variables. The coordinates of ${\cal N}=(1,1)$ superspace are $(x^{\alpha  \dot{\alpha }},~\theta^{\alpha },~\overline{\theta }^{\dot{\alpha }})$. The hermitian conjugation are defined as $(\theta ^{\alpha })^*=i~ \theta _{\alpha }$, \quad
$(\overline{\theta }^{\dot{\alpha }})^*=i~ \overline{\theta }_{\dot{\alpha }}$, \quad
$(x^{\alpha  \dot{\alpha }})^*=-i~ x_{\alpha  \dot{\alpha }}$.
The BFNC superspace is defined as,
\begin{eqnarray}
\label{BFNC_algebra}
\left[x^{\alpha  \dot{\alpha }},~\theta ^{\beta }\right]=\frac{i}{2}~\Lambda ^{\alpha  \dot{\alpha } \beta },
\end{eqnarray}
where $\Lambda ^{\alpha  \dot{\alpha } \beta }$ denotes a BFNC parameter. The other commutators and anti-commutators that do not appear in the above are vanishing. This algebra is consistent in euclidean signature. The chiral and anti chiral sectors are independent.
BFNC star product of two superfields ${\bf F}$ and ${\bf G}$ is,
\begin{eqnarray}
\label{star_product_expansion}
{\bf F} * {\bf G}&=&{\bf F}~ {\bf G}-\frac{i}{2}~\Lambda ^{k \alpha }\left(\partial _{\alpha }{\bf F}\right)\left(\partial _k{\bf G}\right)+(-1)^{|{\bf F}|}~\frac{i}{2}~\Lambda ^{k \alpha }\left(\partial _k{\bf F}\right)\left(\partial _{\alpha }{\bf G}\right) \nonumber \\
&&+\frac{1}{8}~\Lambda ^{k \alpha }~\Lambda ^{l \beta }\left(\partial _k\partial _l{\bf F}\right)\left(\partial _{\alpha }\partial _{\beta }{\bf G}\right)+\frac{1}{8}~\Lambda ^{k \alpha }~\Lambda ^{l \beta }\left(\partial _{\alpha }\partial _{\beta }{\bf F}\right)\left(\partial _k\partial _l{\bf G}\right) \nonumber \\
&&+(-1)^{|{\bf F}|}~\frac{1}{4}~\Lambda ^{k \alpha }~\Lambda ^{l \beta }\left(\partial _{\beta }\partial _k{\bf F}\right)\left(\partial _{\alpha }\partial _l{\bf G}\right)\nonumber \\
&&-\frac{i}{16}~\Lambda ^{k \alpha }~\Lambda ^{l \beta }~\Lambda ^{m \zeta }\left(\partial _{\alpha }\partial _l\partial _m{\bf F}\right)\left(\partial _{\beta }\partial _{\zeta }\partial _k{\bf G}\right)\nonumber \\
&&+(-1)^{|{\bf F}|}~\frac{i}{16}~\Lambda ^{k \alpha }~\Lambda ^{l \beta }~\Lambda ^{m \zeta }\left(\partial _{\alpha }\partial _{\beta }\partial _m{\bf F}\right)\left(\partial _{\zeta }\partial _k\partial _l{\bf G}\right)\nonumber \\
&&-\frac{1}{64}~\Lambda ^{k \alpha }~\Lambda ^{l \beta }~\Lambda ^{m \zeta }~\Lambda ^{n \iota }\left(\partial _{\alpha }\partial _{\zeta }\partial _l\partial _n{\bf F}\right)\left(\partial _{\beta }\partial _{\iota }\partial _k\partial _m{\bf G}\right),
\end{eqnarray}
where $\Lambda ^{k \beta }~\partial _k= \frac{1}{2}~\Lambda ^{\alpha  \dot{\alpha } \beta }~\partial _{\alpha  \dot{\alpha }}$, \quad $|{\bf F}|$ is the grade of ${\bf F}$ that equals zero  for a Bosonic element and one for a Fermionic element.

\section{Review of super Yang-Mills  model  in NAC superspace}
\label{review}

In this section we review the work of \cite{Grisaru:2005we}. They define super Yang-Mills theories in NAC superspace and extend background field method in NAC superspace. In NAC superspace we can define covariant derivatives by using scalar prepotential V, where $V$ is in the adjoint representation of gauge group. The supergauge transformation is defined as,
\begin{eqnarray}
\label{}
e^V_*\to e^{V'}_* = e^{i ~\overline{\Lambda }}_* *e^V_* * e^{-i~ \Lambda }_*,
\end{eqnarray}
where $\Lambda$ and $\overline{\Lambda }$ are chiral and antichiral superfields. In gauge antichiral representation, covariant derivatives is,
\begin{eqnarray}
\label{}
\overline\nabla _A&\equiv&\left(\overline\nabla _{\alpha },\quad \overline\nabla _{\dot{\alpha }},\quad \overline\nabla _{\alpha  \dot{\alpha }}\right)=\left(D_{\alpha },\quad e^{V}_* *\overline{D}_{\dot{\alpha }}~e^{-V}_*,\quad -i~\left\{\overline\nabla _{\alpha },~\overline\nabla _{\dot{\alpha }}\right\}_*\right),
\end{eqnarray}
where V is pure imaginary $V^{\dag}=-V$. From the definition of covariant derivatives, we can define connections $\overline\nabla _A=\overline{D}_A-i ~\overline\Gamma _A$, where
\begin{eqnarray}
\label{}
\overline{\Gamma }_{\alpha }=0,\quad \overline{\Gamma }_{\dot{\alpha }}=i~ e^V_**\overline{D}_{\dot{\alpha }}~e^{-V}_*,\quad \overline{\Gamma }_{\alpha  \dot{\alpha }}=-i ~D_{\alpha }~\overline{\Gamma }_{\dot{\alpha }}.
\end{eqnarray}
By using covariant derivatives we can define superfield strengths,
\begin{eqnarray}
\label{}
\widetilde{W}_{\alpha }=-\frac{1}{2}\left[\overline\nabla ^{\dot{\alpha }},~\overline\nabla _{\alpha  \dot{\alpha }}\right]_*,\quad \overline{W}_{\dot{\alpha }}=-\frac{1}{2}\left[\overline\nabla ^{\alpha },~\overline\nabla _{\alpha  \dot{\alpha }}\right]_*.
\end{eqnarray}
By using the Bianchi's identies of covariant derivatives, we can obtain $\overline\nabla ^{\alpha }*\widetilde{W}_{\alpha }+\overline\nabla ^{\dot{\alpha }}*\overline{W}_{\dot{\alpha }}=0$. The infinitesimal supergauge transformations is,
\begin{eqnarray}
\label{}
\delta~  \widetilde{W}_{\alpha }&=&i\left[\overline{\Lambda },~\widetilde{W}_{\alpha }\right]_*,\quad \delta~  \overline{W}_{\dot{\alpha }}=i\left[\overline{\Lambda },~\overline{W}_{\dot{\alpha }}\right]_*.
\end{eqnarray}
Accroding to background field method,
\begin{eqnarray}
\label{}
\nabla _{\alpha }&=&\boldnabla _{\alpha }=D_{\alpha },\quad \nabla _{\dot{\alpha }}=e^V_**\boldnabla _{\dot{\alpha }}*e^{-V}_*=e^V_**e^U_**\overline{D}_{\dot{\alpha }}~e^{-U}_**e^{-V}_*,\nonumber\\
\overline{\Phi }&=&\bold{\overline{\Phi }},\quad \Phi =e^V_**\bold{\Phi} *e^{-V}_*=e^V_**e^U_**\phi *e^{-U}_**e^{-V}_*,
\end{eqnarray}
where $U$ is the background prepotential. Under quantum transformations,
\begin{eqnarray}
\label{quantum transformations}
e^V_*&\to&  e^{i ~\overline{\Lambda }}_**e^V_**e^{-i~ \Lambda }_*,\quad e^U_*\to e^U_*,\nonumber\\
\nabla _A&\to&  e^{i ~\overline{\Lambda }}_**\nabla _A*e^{-i~ \overline{\Lambda }}_*,\quad \boldnabla _A\to \boldnabla _A,\nonumber\\
\bold{\Phi}&\to&  e^{i ~\Lambda }_**\bold{\Phi}*e^{-i ~\Lambda }_*,\quad\overline{\bold{\Phi}}\to  e^{i ~\overline{\Lambda }}_**\bold{\overline{\Phi }}*e^{-i ~\overline{\Lambda }}_*,
\end{eqnarray}
where $\boldnabla _{\alpha }~\overline{\Lambda }=\boldnabla _{\dot{\alpha }}~\Lambda =0$. Under background transformations,
\begin{eqnarray}
\label{background transformations}
e^V_*&\to&  e^{i~ \overline{\lambda }}_**e^V_**e^{-i~ \overline{\lambda }}_*,\quad e^U_*\to e^{i~ \overline{\lambda }}_**e^U_**e^{-i ~\lambda}_*,\nonumber\\
\nabla _A&\to&  e^{i ~\overline{\lambda }}_**\nabla _A *e^{-i ~\overline{\lambda }}_*,\quad \boldnabla _A\to e^{i ~\overline{\lambda }}_**\boldnabla _A *e^{-i~ \overline{\lambda }}_*,\nonumber\\
\bold{\Phi}&\to&  e^{i~ \overline{\lambda }}_**\bold{\Phi}*e^{-i ~\overline{\lambda }}_*,\quad\overline{\bold{\Phi}}\to  e^{i ~\overline{\lambda }}_**\bold{\overline{\Phi }}*e^{-i~ \overline{\lambda }}_*,
\end{eqnarray}
where $\overline{D}_{\dot{\alpha }}~\lambda =D_{\alpha }~\overline{\lambda }=0$. We can define invariant action under transformation eq.~(\ref{quantum transformations}) and eq.~(\ref{background transformations}),
\begin{eqnarray}
\label{}
S&=&\frac{1}{2g^2}\int d^4x ~d^2\overline{\theta }{\rm ~Tr}\left(\overline{W}^{\dot{\alpha }}~\overline{W}_{\dot{\alpha }}\right)+\int d^4x ~d^4\theta  {\rm ~Tr}\left(\overline{\Phi } *\Phi \right)\nonumber\\
&&-\frac{1}{2}~m\int d^4x~ d^2\theta  {\rm ~Tr}\left(\Phi ^2\right)-\frac{1}{2}~\overline{m}\int d^4x ~d^2\overline{\theta } {\rm ~Tr}\left(\overline{\Phi }^2\right).
\end{eqnarray}
The gauge--fixing functions is $f = \overline{\boldnabla}^2 * V$, $\overline{f} = {\boldnabla}^2 * V$, we obtain the total action $S_{\rm tot} = S + S_{\rm GF} + S_{\rm gh} $, where $S_{\rm gh}$ contains background covariantly chiral and antichiral FP and NK ghost superfields,
\begin{eqnarray}
\label{}
S_{\rm gh} =  \int d^4x~ d^4 \theta ~ \left[ \overline{c}' ~c - c'~\overline{c} + .....+ \overline{b}~ b \right],
\end{eqnarray}
The pure gauge part of action corresponding to one-loop calculation is,
\begin{eqnarray}
S_{\rm inv} + S_{\rm GF} = -\frac{1}{2g^2} \int d^4x ~d^4 \theta~ {\rm Tr}\Big[ (e^{V}_* *{\overline{\boldnabla}}^{\dot{\alpha}} * e^{-V}_*)* D^2 ~(e^{V}_* *\overline{\boldnabla}_{\dot{\alpha}} * e^{-V}_*) \nonumber\\
~~~~~~~~~~~~~~~~~~~~+\frac{1}{\alpha}~ V * (\overline{\boldnabla}^2~ D^2 + D^2~ \overline{\boldnabla}^2) * V \Big].
\end{eqnarray}
To deal with action for  full (not background) covariantly chiral matter,
\begin{eqnarray}
\label{}
S_{\rm mat} &=&\int d^4x ~d^4\theta  {\rm ~Tr}\left(\overline{\Phi } *\Phi \right)=\int d^4x ~d^4 \theta  {\rm ~Tr}\left(\overline{\bold{\Phi}}* e^V_* * \bold{\Phi}* ~e^{-V}_*\right) \nonumber\\
&=&\int d^4x ~d^4 \theta {\rm ~Tr}\Big\{ \overline{\bold{\Phi}}* \bold{\Phi}+\overline{\bold{\Phi}}* [V,~\bold{\Phi}]_*+\cdots\Big\},
\end{eqnarray}
we can write down a formal effective interaction lagrangian,
\begin{eqnarray}
\label{}
\overline{S}_0+\overline{S}_1+\overline{S}_2=\int d^4x~ d^2\theta  {\rm ~Tr}\Bigg\{\overline{\xi }\left(\square _0-m~ \overline{m}\right)\xi +\frac{1}{2}~\overline{\xi }~D^2\left(\overline{\nabla }^2-\overline{D}^2\right)\xi +\frac{1}{2}~\overline{\xi }\left(\square _--\square _0\right)\xi \Bigg\},
\end{eqnarray}
where $\xi ~,~ \overline{\xi}$ are unconstrained quantum fields. In a one-loop diagram, the first vertex must appear once and only once. In order to do calculation in superspace we will expand star product at the end. In momentum superspace, the propagator for quantum gauge superfields is,
\begin{eqnarray}
\label{}
{\langle} V^A(1)~V^B(2){\rangle}=-g^2~\delta ^{AB}~\frac{1}{p_1{}^2}~\delta ^{(8)}\left(\Pi _1+\Pi _2\right),
\end{eqnarray}
the vertices are,
\begin{eqnarray}
&& S_{\rm inv} + S_{\rm GF} \rightarrow \frac{1}{2 g^2} \int {\cal P}^{ABC}(\pi_2, ~\pi_3) \Big\{V^A(1)~\overline{\Gamma}^{\alpha \dot{\alpha} B}(2)~ p_{3 \alpha \dot{\alpha}}~ V^C(3)\nonumber \\
&&~~~~+\frac{1}{2}~V^A(1)~p_{2 \alpha \dot{\alpha}}~\overline{\Gamma}^{\alpha \dot{\alpha}B}(2)~V^C(3) +i~V^A(1)~\overline{W}^{\dot{\alpha}B}(2)~\left(\widetilde{\overline{D}}_{\dot{\alpha}}~V^C(3) \right)\nonumber \\
&&~~~~+i~V^A(1)~\widetilde{W}^{\alpha B}(2)~\left(\widetilde{D}_{\alpha} ~V^C(3) \right)\Big\}\nonumber \\
&& + \frac{1}{2 g^2} \int {\cal Q}^{ABCD}(\pi_1,~ \pi_2, ~\pi_3, ~\pi_4) \Big\{ \frac{1}{2}~V^A(1) ~\overline{\Gamma}^{\alpha \dot{\alpha}B}(2) ~\overline{\Gamma}_{\alpha \dot{\alpha}}^C (3)~V^D(4)\nonumber \\
&&~~~~~~~~~~~~~+ V^A(1)~\overline{W}^{\dot{\alpha} B}(2) ~\overline{\Gamma}_{\dot{\alpha} }^C (3)~V^D(4) \Big\}.
\end{eqnarray}
The propagator for quantum (anti)chiral superfields is,
\begin{eqnarray}
\label{}
{\langle} \overline{\xi }^A(1)~\xi ^B(2){\rangle}=\delta ^{AB}~\frac{1}{p_1{}^2+m ~\overline{m}}~\delta ^{(8)}\left(\Pi _1+\Pi _2\right),
\end{eqnarray}
the vertices are,
\begin{eqnarray}
&& \overline{S}_1 = \frac{i}{4} \int {\cal P}^{ABC}(\pi_1, ~\pi_2)~\overline{\Gamma}^{\dot{\alpha} A}(1) ~\big[~ \xi^{B}(2)~\widetilde{\overline{D}}_{\dot{\alpha}}~\widetilde{D}^2~\overline{\xi}^{C}(3) - \widetilde{\overline{D}}_{\dot{\alpha}}~\xi^{B}(2)~\widetilde{D}^2~ \overline{\xi}^C(3) ~\big],\nonumber \\
&&\overline{S}_2 = \frac{1}{4} \int {\cal P}^{ABC}(\pi_1, ~\pi_2) ~\big[ - p_{2 \alpha\dot{\alpha}} + p_{3 \alpha \dot{\alpha}}\big]~\overline{\Gamma}^{\alpha \dot{\alpha} A}(1) ~\xi^{B}(2) ~\overline{\xi}^{C}(3),\nonumber \\
&& \overline{S}'_2 = \frac{i}{4} \int {\cal P}^{ABC}(\pi_1, ~\pi_2)~\overline{W}^{\dot{\alpha} A}(1) ~\big[~ \xi^{B}(2)~\widetilde{\overline{D}}_{\dot{\alpha}} ~\overline{\xi}^{C}(3) -\widetilde{\overline{D}}_{\dot{\alpha}}~\xi^{B}(2) ~ \overline{\xi}^C(3) ~\big],\nonumber \\
&& \overline{S}'_1= - \frac{1}{4}\int {\cal Q}^{ABCD} (\pi_1, ~\pi_2, ~\pi_3, ~\pi_4)~\xi^A(1)~\overline{\Gamma}^{\dot{\alpha} B }(2)~\overline{\Gamma}_{\dot{\alpha}}^C(3)~\widetilde{D}^2 ~\overline{\xi}^D(4),\nonumber \\
&& \overline{S}''_2 = - \frac{1}{4}\int {\cal Q}^{ABCD} ( \pi_1,~\pi_2,~ \pi_3,~ \pi_4)~\times\nonumber \\
&&~~~~~~~~~~~~~~~~~~~~~~~~\xi^A(1) \left(\overline{\Gamma}^{\dot{\alpha} B }(2)~\overline{W}_{\dot{\alpha}}^C(3) +\overline{W}^{\dot{\alpha} B }(2)~\overline{\Gamma}_{\dot{\alpha}}^C(3)~\right)~\overline{\xi}^D(4),\nonumber \\
&& \overline{S}'''_2 =  -\frac{1}{4}\int {\cal Q}^{ABCD} ( \pi_1, ~\pi_2, ~\pi_3, ~\pi_4)~\xi^A(1)~\overline{\Gamma}^{\alpha \dot{\alpha} B }(2)~\overline{\Gamma}_{\alpha \dot{\alpha}}^C(3)~\overline{\xi}^D(4).
\end{eqnarray}

\section{Effective action of dermored super Yang-Mills action in BFNC superspace}
\label{effective action}

Because the BFNC star product is similar to the NAC star product, we can use the method in section~\ref{review} to study the renormalization property of deformed super Yang-Mills action in BFNC superspace.  
We only need to replace the NAC star product by BFNC star product, this will only change the definition of ${\cal P}$ and ${\cal Q}$, where $\pi _1\wedge \pi _2=-\frac{i}{2}\left(\Lambda ^{k \alpha }~p_{1k}~\pi _{2\alpha }-\Lambda ^{k \alpha }~p_{2k}~\pi _{1\alpha }\right)$. At the first order of BFNC parameters, only the chiral mater action contribute to the effective action and there are at most three points contributions. The effective action is, 
\begin{eqnarray}
\label{}
\Gamma _{{\rm gauge}}{}^{(1)}&=&\int d^4x~ d^4\theta ~\Bigg\{\frac{1}{8} (N S)~ \overline{\Gamma }^{\dot{{\alpha_{1}}} {c_{1}}} ~\overline{W}_{\dot{{\alpha_{1}}}}^{{c_{1}}}+\frac{1}{8} \left(i \sqrt{N} S\right) \Lambda ^{\dot{{\alpha_{1}}}} ~\overline{\Gamma }_{\dot{{\alpha_{1}}}}^0 ~\overline{{WW}}^{{c_{1}} {c_{1}}}\nonumber\\
&&+\frac{1}{4} \left(i \sqrt{N} S\right) \overline{\theta }^{\dot{{\alpha_{1}}}} ~\overline{\Gamma }^{\dot{{\alpha_{2}}} 0} \left(\overline{\Gamma }^{{\alpha_{3}}}\right)_{\dot{{\alpha_{1}}}}^{{c_{1}}} \left({p\Lambda }_{{\alpha_{3}}} ~\overline{W}_{\dot{{\alpha_{2}}}}^{{c_{1}}}\right)\nonumber\\
&&+\frac{1}{12} \left(-i \sqrt{N} S\right) \overline{\theta }^{\dot{{\alpha_{1}}}}~ \overline{\Gamma }_{\dot{{\alpha_{1}}}}^0 \left({p\Lambda }^{{\alpha_{2}}} ~\overline{\Gamma }_{{\alpha_{2}}}^{\dot{{\alpha_{3}}} {c_{1}}}\right) ~\overline{W}_{\dot{{\alpha_{3}}}}^{{c_{1}}}\nonumber\\
&&+\frac{1}{8} \left(-i \sqrt{N} S\right) \Lambda ^{\dot{{\alpha_{1}}}} ~\overline{\theta }^{\dot{{\alpha_{2}}}} ~\overline{\Gamma }^{\dot{{\alpha_{3}}} 0} \left(\partial^{{\alpha_{4}}}{}_{\dot{{\alpha_{1}}}} ~\overline{\Gamma }_{{\alpha_{4}} \dot{{\alpha_{2}}}}^{{c_{1}}}\right) ~\overline{W}_{\dot{{\alpha_{3}}}}^{{c_{1}}}\nonumber\\
&&+\frac{1}{8} \left(i \sqrt{N} S\right) \Lambda ^{\dot{{\alpha_{1}}}} ~\overline{\theta }_{\dot{{\alpha_{1}}}} ~\overline{\Gamma }^{\dot{{\alpha_{2}}} 0} \left(\partial^{{\alpha_{3}} \dot{{\alpha_{4}}}} ~\overline{\Gamma }_{{\alpha_{3}} \dot{{\alpha_{2}}}}^{{c_{1}}}\right) ~\overline{W}_{\dot{{\alpha_{4}}}}^{{c_{1}}}\nonumber\\
&&+\frac{1}{8} \left(-i \sqrt{N} S\right) \Lambda ^{\dot{{\alpha_{1}}}} ~\overline{\theta }_{\dot{{\alpha_{1}}}} ~\overline{\Gamma }^{\dot{{\alpha_{2}}} 0} \left(\partial^{{\alpha_{3}}}{}_{\dot{{\alpha_{2}}}} ~\overline{\Gamma }_{{\alpha_{3}}}^{\dot{{\alpha_{4}}} {c_{1}}}\right) ~\overline{W}_{\dot{{\alpha_{4}}}}^{{c_{1}}}\nonumber\\
&&+\frac{1}{8} \left(i \sqrt{N} S\right) \Lambda ^{{\alpha_{1}} \dot{{\alpha_{2}}} {\alpha_{3}}}~ \overline{\theta }^{\dot{{\alpha_{4}}}} ~\overline{\Gamma }^{\dot{{\alpha_{5}}} 0} ~\overline{\Gamma }_{{\alpha_{1}} \dot{{\alpha_{2}}}}^{{c_{1}}} \left(\partial_{{\alpha_{3}} \dot{{\alpha_{4}}}} ~\overline{W}_{\dot{{\alpha_{5}}}}^{{c_{1}}}\right)\nonumber\\
&&+\frac{1}{24} \left(i \sqrt{N} S\right) \Lambda ^{{\alpha_{1}} \dot{{\alpha_{2}}} {\alpha_{3}}} ~\overline{\theta }^{\dot{{\alpha_{4}}}} ~\overline{\Gamma }_{\dot{{\alpha_{2}}}}^0 \left(\partial_{{\alpha_{1}} \dot{{\alpha_{4}}}} ~\overline{\Gamma }_{{\alpha_{3}}}^{\dot{{\alpha_{5}}} {c_{1}}}\right) ~\overline{W}_{\dot{{\alpha_{5}}}}^{{c_{1}}}\nonumber\\
&&+\frac{1}{8} \left(-i \sqrt{N} S\right) \Lambda ^{{\alpha_{1}} \dot{{\alpha_{2}}} {\alpha_{3}}} ~\overline{\theta }_{\dot{{\alpha_{2}}}} ~\overline{\Gamma }^{\dot{{\alpha_{4}}} 0} \left(\partial_{{\alpha_{1}}}^{\dot{{\alpha_{5}}}} ~\overline{\Gamma }_{{\alpha_{3}} \dot{{\alpha_{4}}}}^{{c_{1}}}\right) ~\overline{W}_{\dot{{\alpha_{5}}}}^{{c_{1}}}\nonumber\\
&&+\frac{1}{12} \left(i \sqrt{N} S\right) \Lambda ^{{\alpha_{1}} \dot{{\alpha_{2}}} {\alpha_{3}}} ~\overline{\theta }_{\dot{{\alpha_{2}}}} ~\overline{\Gamma }^{\dot{{\alpha_{4}}} 0} \left(\partial_{{\alpha_{1}} \dot{{\alpha_{4}}}} ~\overline{\Gamma }_{{\alpha_{3}}}^{\dot{{\alpha_{5}}} {c_{1}}}\right) ~\overline{W}_{\dot{{\alpha_{5}}}}^{{c_{1}}}\Bigg\}.
\end{eqnarray}
We can verify that $\Gamma _{{\rm gauge}}{}^{(1)}$ is supergauge invariant.

\section{Conclusion and outlook}
\label{conclusion}

By replacing the ordinary products of super Yang-Mills model by BFNC star products, we obtain deformed action in BFNC superspace. The renormalization property of the deformed action have been studied by calculating the one-loop 1PI effective action with background field method at the first order of  BFNC parameters. The effective action have BFNC supergauge invariance. But there are new terms in the effective action, which make the deformed action not renormalizable. This imply that the deformed action need additional terms to be renormalizable. 

This is the first step to construct renormalizable gauge model in Bosonic-Fermionic noncommutative (BFNC) superspace. To find the other terms we can add the effective action to the deformed action, then calculate the effective action of the new action. We should repeat this process for several times until we verify that there are not any new terms in the effective action.

\section*{Acknowledgments}
The author would like to thank Professors Li-Ming Cao, Dao-Neng Gao, Min-Xin Huang, MingZhe Li, JianXin Lu, Yan-Gang Miao, Zhi-Guang Xiao for helpful discussions.

\newpage

\appendix
\setcounter{equation}{0}
\renewcommand\theequation{A\arabic{equation}}

\section{Conventions}

Gates's conventions:
\begin{eqnarray}
\label{}
C_{\alpha  \beta }=C_{\dot{\alpha } \dot{\beta }}=\left(
\begin{array}{cc}
 0 & -i \\
 i & 0
\end{array}
\right),\quad
C^{\alpha  \beta }=C^{\dot{\alpha } \dot{\beta }}=\left(
\begin{array}{cc}
 0 & i \\
 -i & 0
\end{array}
\right).
\end{eqnarray}
The index are raised and lowed by $C$:\quad
$\psi ^{\alpha }=C^{\alpha \beta }~\psi _{\beta },\quad \psi _{\alpha }=\psi ^{\beta }~C_{\beta \alpha }$.

Wess's conventions:
\begin{eqnarray}
\label{}
\epsilon _{\alpha  \beta }=\epsilon _{\dot{\alpha } \dot{\beta }}=\left(
\begin{array}{cc}
 0 & -1 \\
 1 & 0
\end{array}
\right),\quad \epsilon ^{\alpha  \beta }=\epsilon ^{\dot{\alpha } \dot{\beta }}=\left(
\begin{array}{cc}
 0 & 1 \\
 -1 & 0
\end{array}
\right).
\end{eqnarray}
The index are raised and lowed by $\epsilon$:\quad
$\psi ^{\alpha }=\epsilon ^{\alpha \beta }~\psi _{\beta },\quad \psi _{\alpha }=\epsilon _{\alpha \beta }~\psi ^{\beta }$.\quad
$\sigma ^k{}_{\beta  \dot{\beta }}$ is Pauli matrix,
\begin{eqnarray}
\label{}
\sigma ^0=\left(
\begin{array}{cc}
 -1 & 0 \\
 0 & -1
\end{array}
\right),\quad \sigma ^1=\left(
\begin{array}{cc}
 0 & 1 \\
 1 & 0
\end{array}
\right),\quad \sigma ^2=\left(
\begin{array}{cc}
 0 & -i \\
 i & 0
\end{array}
\right),\quad \sigma ^3=\left(
\begin{array}{cc}
 1 & 0 \\
 0 & -1
\end{array}
\right).
\end{eqnarray}
Define:\quad
$\overline{\sigma }^{k \dot{\alpha }\alpha }=\epsilon ^{\dot{\alpha }\dot{\beta }}~\epsilon ^{\alpha \beta }~\sigma ^k{}_{\beta  \dot{\beta }}$.\quad
We have the following relations:\quad
$\sigma ^k{}_{\alpha  \dot{\alpha }}~\overline{\sigma }_k{}^{\dot{\beta }\beta }=-2~ \delta _{\alpha }{}^{\beta }~\delta _{\dot{\alpha }}{}^{\dot{\beta }}$,\quad 
$\sigma ^k{}_{\alpha  \dot{\alpha }}~\sigma _{k \beta \dot{\beta }}=-2~\epsilon _{\alpha \beta }~\epsilon _{\dot{\alpha }\dot{\beta }}$,\quad 
$\sigma ^k{}_{\alpha  \dot{\beta }}~\overline{\sigma }^{l \dot{\beta }\alpha }=-2~\eta ^{{kl}}$.

The relation between the convention of Gates's and Wess's,

\begin{itemize}
\label{}
\item
$C_{\alpha \beta }=i ~\epsilon _{\alpha \beta }$,\quad 
$C^{\alpha \beta }=i~ \epsilon ^{\alpha  \beta }$,\quad 
$C_{\dot{\alpha }\dot{\beta }}=i~ \epsilon _{\dot{\alpha }\dot{\beta }}$,\quad 
$C^{\dot{\alpha }\dot{\beta }}=i ~\epsilon ^{\dot{\alpha } \dot{\beta }}$.

\item
Definition:\quad
$\psi ^{\alpha (G)}=\psi ^{\alpha (W)}$,\quad 
$q_k{}^{(G)}=q_k{}^{(W)}$,\quad 
$\sigma ^{k \alpha  \dot{\beta }(G)}=\overline{\sigma }^{k\dot{\beta }\alpha  (W)}$.

\item
$\psi _{\alpha }{}^{(G)}=-i ~\psi _{\alpha }{}^{(W)}$.\quad
Proof:\quad
$\psi _{\alpha }{}^{(G)}=\psi ^{\beta (G)}~C_{\beta \alpha }=i~ \psi ^{\beta (W)}~\epsilon _{\beta  \alpha }=-i ~\psi _{\alpha }{}^{(W)}$.

\item
Identity:\quad
$\sigma ^k{}_{\alpha  \dot{\beta }}{}^{(G)}=-\sigma ^k{}_{\alpha  \dot{\beta }}{}^{(W)}$.

\item
Definition:\quad
$q^{2 (G)}=\frac{1}{2}~q^{\alpha  \dot{\beta }(G)}~q_{\alpha  \dot{\beta }}{}^{(G)}$,\quad
$q_{\alpha  \dot{\beta }}{}^{(G)}=\sigma ^k{}_{\alpha  \dot{\beta }}{}^{(G)}~q_k{}^{(G)}$,\quad
$q^{\alpha  \dot{\beta }(G)}=\sigma ^{k \alpha  \dot{\beta }(G)}~q_k{}^{(G)}$.

\item
Identity:\quad
$q_{\alpha  \dot{\beta }}{}^{(G)}=-\sigma ^k{}_{\alpha  \dot{\beta }}{}^{(W)}~q_k{}^{(W)}$,\quad
$q^{\alpha  \dot{\beta }(G)}=\overline{\sigma } ^{k \dot{\beta }\alpha  (W)}~q_k{}^{(W)}$.
\end{itemize}

We use the following definition in this paper:\quad
$q^{\alpha  \dot{\beta }}=q^{\alpha  \dot{\beta }(G)}$,\quad 
$q_{\alpha  \dot{\beta }}=q_{\alpha  \dot{\beta }}{}^{(G)}$,\quad 
$q_k=q_k{}^{(G)}$,\quad
$\overline{\sigma }^{k\dot{\beta }\alpha }=\overline{\sigma }^{k\dot{\beta }\alpha (W)}$,\quad 
$\sigma ^k{}_{\alpha  \dot{\beta }}=\sigma ^k{}_{\alpha  \dot{\beta }}{}^{(W)}$.

Identity:\quad
$\frac{1}{2}~q^{\alpha  \dot{\beta }}~q_{\alpha  \dot{\beta }}=q^k~q_k$.\quad
Proof:\quad
$\frac{1}{2}~q^{\alpha  \dot{\beta }}~q_{\alpha  \dot{\beta }}=-\frac{1}{2}~\overline{\sigma }_k{}^{\dot{\beta }\alpha }~q^k~\sigma ^l{}_{\alpha  \dot{\beta }}~q_l=-\frac{1}{2}~(-2)~q^k~q_l~\delta _k{}^l=q^k~q_k$.

\section{Hermitian conjugation}
\begin{itemize}
\label{}
\item
The hermitian conjugation $*$ will give $-1$ if we apply it on spinorial coordinates twice, and will give $1$ if we apply it on bosonic coordinates twice.
If we define $\left\{\theta ^{\alpha },~\theta ^{\beta }\right\}=F^{\alpha \beta }$ then we find:\quad 
$\left\{\theta _{\alpha },~\theta _{\beta }\right\}=F_{\alpha \beta }$.
From $\left(\theta ^{\alpha }\right)^*=i ~\theta _{\alpha }$ we have:\quad 
$\left(\theta _{\alpha }\right){}^*=-i ~\theta ^{\alpha }$, \quad 
$\left(F^{\alpha \beta }\right)^*=-F_{\alpha \beta }$, \quad
$\left(F_{\alpha \beta }\right){}^*=-F^{\alpha \beta }$.

From definition
$({A~B})^*=B^*~A^*$,\quad
$[A,~B]={A~B}-{B~A}$,\quad
$\{A,~B\}={A~B}+{B~A}$
we obtain
$[A,~B]^*=-\left[A^*,~B^*\right]$, \quad 
$\{A,~B\}^*=\left\{A^*,~B^*\right\}$.

\item
The hermitian conjugation relation for BFNC parameters:\quad
$\left(x^{\alpha  \dot{\beta }}\right)^*=-x_{\alpha  \dot{\beta }}$,\quad 
$\left(\theta ^{\gamma }\right)^*=i ~\theta _{\gamma }$,\quad 
$\left(\Lambda ^{\alpha  \dot{\beta }\gamma }\right)^*=(-i) ~\Lambda _{\alpha  \dot{\beta }\gamma }$.\quad 
Proof: 
Apply hermitian conjugation on
$\left[x^{\alpha  \dot{\beta }},~\theta ^{\gamma }\right]=\frac{i}{2}~\Lambda ^{\alpha  \dot{\beta }\gamma }$
we obtain
$(-1)\left[\left(x^{\alpha  \dot{\beta }}\right)^*,~\left(\theta ^{\gamma }\right)^*\right]=-\frac{i}{2}~\left(\Lambda ^{\alpha  \dot{\beta }\gamma }\right)^*$,
then we have
$\left[x_{\alpha  \dot{\beta }},~ \theta _{\gamma }\right]=\frac{i}{2}~\Lambda _{\alpha  \dot{\beta }\gamma }$.

\end{itemize}

\section{Transform from configuration superspace to momentum superspace}

\begin{itemize}
\label{}
\item
$\pi _{\alpha }\to  i~ \partial _{\alpha }$,\quad 
$\overline{\pi }_{\dot{\alpha }}\to  i~ \overline{\partial }_{\dot{\alpha }}$,\quad 
$p_{\alpha  \dot{\beta }}\to  i ~\partial _{\alpha  \dot{\beta }}$,\quad
$p_k\to  i~\partial _k$,\quad 
$p_k\to  \frac{i}{2}~\overline{\sigma }_k{}^{\dot{\beta }\alpha }~\partial _{\alpha  \dot{\beta }}$,\quad
${\Lambda p}^{\alpha }\to  i~ \Lambda ^{k \alpha }~\partial _k$,\quad 
${\Lambda p}^{\alpha }\to  \frac{i}{2}~\Lambda ^{k \alpha }~\overline{\sigma }_k{}^{\dot{\beta }\gamma }~\partial _{\gamma  \dot{\beta }}$,\quad 
${\Lambda p}_{\alpha }\to \Lambda ^k{}_{\alpha }~\partial _k$.

\item
${\Lambda p}^{\alpha }\to  i~ {p\Lambda }^{\alpha },\quad {\Lambda p}_{\alpha }\to i ~{p\Lambda }_{\alpha },\quad {\Lambda p}^2\to -{p\Lambda }^2$.\quad
Proof:\quad
${\Lambda p}^{\alpha }=\Lambda ^{{k\alpha }}~p_k\to  \Lambda ^{{k\alpha }}~i~ \partial _k=i ~{p\Lambda }^{\alpha }$,\quad
${\Lambda p}_{\alpha }=(-i) ~\Lambda ^k{}_{\alpha }~p_k\to  (-i)~\Lambda ^k{}_{\alpha }~(i)~\partial _k=i~ {p\Lambda }_{\alpha }$,\quad
${\Lambda p}^2=\frac{1}{2}~{\Lambda p}^{\alpha }~{\Lambda p}_{\alpha }\to  \frac{1}{2}~i~ {p\Lambda }^{\alpha }~i~ {p\Lambda }_{\alpha }=-{p\Lambda }^2$.

\item
Definition:\quad
$\widetilde{D}^2=-\pi ^2-\pi ^{\alpha }~\overline{\eth}^{\dot{\beta }}~p_{\alpha  \dot{\beta }}-\overline{\eth}^2~p^2$.

\item
$\overline{D}_{\dot{\alpha }}\to  -i~\overline{\pi }_{\dot{\alpha }}$.\quad
Proof:\quad
$\overline{D}_{\dot{\alpha }}=\overline{\partial }_{\dot{\alpha }}$,\quad 
$\overline{\partial }_{\dot{\alpha }}\to  -i ~\overline{\pi }_{\dot{\alpha }}$.

\item
Expand $\tilde{D}^2$,\quad
$\tilde{D}^2\left(\Pi _{a_1}\right)~\delta \left(\Pi _{a_1}+\Pi _{a_2}\right)=\int\tilde{D}^2\left(\Pi _{a_1}, ~Z_{b_1}\right)~e^{i\left(\Pi _{a_1}+\Pi _{a_2}\right)Z_{b_1}}~{dZ}_{b_1}$,\\
where
$\tilde{D}^2\left(\Pi _a, ~Z_b\right)=-\pi ^2\left(\Pi _a\right)-i ~\pi ^{\alpha }\left(\Pi _a\right)~\overline{\theta }^{\dot{\beta }}~p_{\alpha  \dot{\beta }}\left(\Pi _a\right)+\overline{\theta }^2~p^2\left(\Pi _a\right)$.

\end{itemize}

\section{Lie algebra structure constant}

Definition:\quad
$F^B{}_{{AC}}=f^{{ABC}}$,\quad $D^B{}_{{AC}}=d^{{ABC}}$.

We have the following relations,
\begin{itemize}
\label{}
\item
Transpose: \quad
${\rm t}\left(F^A\right)=-F^A, \quad {\rm t}\left(D^A\right)=D^A$.

\item
Trace:\quad
${\rm ~tr~}({X~Y})={\rm ~tr~}({Y~X})$.

\item
Evaluate trace of F and D,
\begin{eqnarray}
\label{}
{\rm ~tr~} F^{{A_1}} D^{{A_2}} D^{{A_3}}&=& \frac{N}{2 \sqrt{2}} ~f^{{A_1} {A_2} {A_3}} ~c_{{A_1}},\quad
{\rm ~tr~} F^{{A_1}} F^{{A_2}} F^{{A_3}}= -\frac{N}{2 \sqrt{2}} ~f^{{A_1} {A_2} {A_3}} ~c_{{A_3}},\nonumber\\
{\rm ~tr~} F^{{A_1}} F^{{A_2}} D^{{A_3}}&=& -\frac{N}{2 \sqrt{2}}~ d^{{A_1} {A_2} {A_3}} ~c_{{A_1}}-\frac{N }{2 \sqrt{2}}~d^{{A_1} {A_2} {A_3}} ~c_{{A_2}}+\frac{N }{2 \sqrt{2}}~d^{{A_1} {A_2} {A_3}} ~c_{{A_3}},\nonumber\\
{\rm ~tr~} D^{{A_1}} D^{{A_2}} D^{{A_3}}&=& -\frac{N}{2 \sqrt{2}} ~d^{{A_1} {A_2} {A_3}} ~c_{{A_1}}-\frac{N}{2 \sqrt{2}} ~d^{{A_1} {A_2} {A_3}}~ c_{{A_2}}+\frac{N}{2 \sqrt{2}} ~d^{{A_1} {A_2} {A_3}} ~c_{{A_3}}\nonumber\\
&&+\frac{N}{\sqrt{2}} ~d^{{A_1} {A_2} {A_3}} ~d_{{A_3}}.
\end{eqnarray}

\item
$N {\rm ~Tr}~(\Gamma ~ W)-{\rm ~Tr}~(\Gamma ){\rm ~Tr}~(W)=N~ \Gamma ^a~W^a$.\quad
Proof:\quad
$N {\rm ~Tr~}(\Gamma ~ W)=N~ \Gamma ^A~W^B{\rm ~Tr~}\left(T^A~T^B\right)=N ~\Gamma ^A~W^B~\delta ^{A B}=N~ \Gamma ^A~W^A=N~\left(\Gamma ^a~W^a+\Gamma ^0~W^0\right)$,\\
${\rm ~Tr~}(\Gamma ){\rm ~Tr~}(W)={\rm ~Tr~}\left(\Gamma ^A~T^A\right){\rm ~Tr~}\left(W^B~T^B\right)=\Gamma ^A~W^B{\rm ~Tr~}\left(T^A\right){\rm ~Tr~}\left(T^B\right)\\=\Gamma ^A~W^B~\sqrt{N}~\delta ^{0A}~\sqrt{N}~\delta ^{0B}=N~ \Gamma ^A~W^B~\delta ^{0A}~\delta ^{0B}=N~\Gamma ^0~W^0$.

\end{itemize}

\section{PQ factor}

There are noncommutative parameters $\Lambda ^{{k\alpha }}$ in $P$ $Q$ factors, at the first order of noncommutative parameters, we only need to expand $P$ $Q$ to the first order of $\Lambda ^{{k\alpha }}$,
\begin{eqnarray}
\label{}
P^{{A_1} {A_2} {A_3}}(x,y)&=&i ~f^{{A_1} {A_2} {A_3}}-\frac{i}{2} ~ d^{{A_1} {A_2} {A_3}} ~{\Lambda p}^{{\alpha_{1}}}(x) ~\pi _{{\alpha_{1}}}(y)+\frac{i}{2}~  d^{{A_1} {A_2} {A_3}}~ {\Lambda p}^{{\alpha_{1}}}(y)~ \pi _{{\alpha_{1}}}(x),\nonumber\\
Q^{{A_1} {A_2} {A_3} {A_4}}(w,x,y,z)&=&-f^{{A_1} {A_2} {A_{01}}}~ f^{{A_3} {A_4} {A_{01}}}+\frac{1}{2} ~d^{{A_{01}} {A_1} {A_2}}~ f^{{A_3} {A_4} {A_{01}}}~ {\Lambda p}^{{\alpha_{1}}}(w) ~\pi _{{\alpha_{1}}}(x)\nonumber\\
&&-\frac{1}{2}~ d^{{A_{01}} {A_1} {A_2}} ~f^{{A_3} {A_4} {A_{01}}} ~{\Lambda p}^{{\alpha_{1}}}(x) ~\pi _{{\alpha_{1}}}(w)\nonumber\\
&&+\frac{1}{2} ~d^{{A_{01}} {A_3} {A_4}} ~f^{{A_1} {A_2} {A_{01}}} ~{\Lambda p}^{{\alpha_{1}}}(y)~ \pi _{{\alpha_{1}}}(z)\nonumber\\
&&-\frac{1}{2}~ d^{{A_{01}} {A_3} {A_4}} ~f^{{A_1} {A_2} {A_{01}}} ~{\Lambda p}^{{\alpha_{1}}}(z) ~\pi _{{\alpha_{1}}}(y).
\end{eqnarray}

\section{Algebraic relations related to $\sigma$}

Definition:\quad
${\Lambda p}^{\alpha }=\Lambda ^{{k\alpha }}~p_k$,\quad 
$p_{\alpha  \dot{\beta }}=-\sigma ^k{}_{\alpha  \dot{\beta }}~p_k$,\quad 
$x^{\alpha  \dot{\alpha }}=\frac{1}{2}~\sigma _k{}^{\alpha  \dot{\alpha }}~x^k$, \quad
$\partial _{\alpha  \dot{\alpha }}~x^{\beta  \dot{\beta }}=\delta _{\alpha }{}^{\beta }~\delta _{\dot{\alpha }}{}^{\dot{\beta }}$,\quad
$\Lambda ^{{kl}}=\Lambda ^{{k\alpha }}~\Lambda ^l{}_{\alpha }$,\quad
$\Lambda ^{\alpha  \dot{\beta }\gamma }=\Lambda ^{k \gamma }~\overline{\sigma }_k{}^{\dot{\beta }\alpha }$,\quad 
$\Lambda _{\alpha  \dot{\beta }}{}^{\gamma }=\Lambda ^{\rho  \dot{\zeta }\gamma }~C_{\rho \alpha }~C_{\dot{\zeta }\dot{\beta }}$,\quad
${p\Lambda }^{\alpha }=\Lambda ^{{k\alpha }}~\partial _k$,\quad 
${p\Lambda }_{\alpha }={p\Lambda }^{\beta }~C_{\beta \alpha }=(-i)~\Lambda ^k{}_{\alpha }~\partial _k$,\quad
$\Lambda ^{\alpha  \dot{\beta }}{}_{\alpha }=\Lambda ^{\dot{\beta }}$,\quad 
$\Lambda ^{\alpha  }{}_{\dot{\beta }\alpha }=\Lambda _{\dot{\beta }}$.

We can prove the following identities.
\begin{itemize}
\label{}
\item
$p_k=\frac{1}{2}~\overline{\sigma }_k{}^{\dot{\beta }\alpha }~p_{\alpha  \dot{\beta }}$.\quad
Proof: \quad
$\overline{\sigma }_l{}^{\dot{\beta }\alpha }~p_{\alpha  \dot{\beta }}=-\overline{\sigma }_l{}^{\dot{\beta }\alpha }~\sigma ^k{}_{\alpha  \dot{\beta }}~p_k=(-1)(-2)~\delta ^k{}_l~p_k=2~p_k$.

\item
$p^{\alpha  \dot{\beta }}=\overline{\sigma }^{k\dot{\beta }\alpha}~p_k$.\quad
Proof:\quad
$p^{\alpha  \dot{\beta }}=C^{\alpha  \gamma }~C^{\dot{\beta }\dot{\rho }}~p_{\gamma  \dot{\rho }}=i~ \epsilon ^{\alpha  \gamma }~i~ \epsilon ^{\dot{\beta }\dot{\rho }}~(-1)~\sigma ^k{}_{\gamma  \dot{\rho }}~p_k=\overline{\sigma }^{k\dot{\beta }\alpha }~p_k$. 
We have used:\quad
$\overline{\sigma }^{m \dot{\alpha }\alpha }=\epsilon ^{\dot{\alpha }\dot{\beta }}~\epsilon ^{\alpha  \beta }~\sigma ^m{}_{\beta  \dot{\beta }},\quad C^{\alpha  \beta }=i ~\epsilon ^{\alpha  \beta }, \quad \psi ^{\alpha }=C^{\alpha  \beta }~\psi _{\beta }$.

\item
$p_{\alpha  \dot{\beta }}~x^{\alpha  \dot{\beta }}=p_k~x^k$,\quad
We have used:\quad
$\sigma ^k{}_{\alpha  \dot{\beta }}~\overline{\sigma }^{l\dot{\beta }\alpha }=(-2)~\eta ^{k l}$.

\item
${\Lambda p}_{\alpha }=(-i)~\Lambda ^k{}_{\alpha }~p_k$.\quad
Proof:\quad
${\Lambda p}_{\alpha }={\Lambda p}^{\beta }~C_{\beta \alpha }=\Lambda ^{{k\beta }}~p_k~i~\epsilon _{\beta \alpha }=(-i)~\Lambda ^k{}_{\alpha }~p_k$,\quad
We have used:\quad
$\Lambda ^k{}_{\alpha }=\epsilon _{\alpha \beta }~\Lambda ^{{k\beta }}$.

\item
$\partial _{\alpha  \dot{\alpha }}=-~\sigma ^k{}_{\alpha  \dot{\alpha }}~\partial _k, \quad
\partial _k=\frac{1}{2}~\overline{\sigma }_k{}^{\dot{\beta }\alpha }~\partial _{\alpha  \dot{\beta }}$.\quad
Proof: \quad
Assume
$\partial _{\alpha  \dot{\alpha }}=a~ \sigma ^k{}_{\alpha  \dot{\alpha }}~\partial _k$,~
apply $\partial _{\alpha  \dot{\alpha }}$ 
on 
$x^{\beta  \dot{\beta }}$ 
we obtain
$\partial _{\alpha  \dot{\alpha }}~x^{\beta  \dot{\beta }}=\frac{1}{2}~\overline{\sigma }_k{}^{\dot{\beta }\beta }~\partial _{\alpha  \dot{\alpha }}~x^k=\frac{1}{2}~a~\overline{\sigma }_k{}^{\dot{\beta }\beta }~\sigma ^l{}_{\alpha  \dot{\alpha }}~\partial _l~x^k=\frac{1}{2}~a~\overline{\sigma }_k{}^{\dot{\beta }\beta }~\sigma ^k{}_{\alpha  \dot{\alpha }}=\frac{1}{2}~a~(-2)~\delta _{\alpha }{}^{\beta }~\delta _{\dot{\alpha }}{}^{\dot{\beta }}$,~
so we have  $a=-1$. 
We have used:\quad 
$\partial _k~x^l=\delta _k{}^l$.

\item
${\Lambda p}^2=\frac{1}{2}~{\Lambda p}^{\alpha }~{\Lambda p}_{\alpha }=-\frac{i}{2}~\Lambda ^{{kl}}~p_k~p_l$.

\item
$\Lambda ^{{\alpha_{1}} \dot{{\alpha_{2}}}{\alpha_{3}} \dot{{\alpha_{4}}}}=\Lambda _{{kl}}~\overline{\sigma }^{k \dot{{\alpha_{2}}}{\alpha_{1}}}~\overline{\sigma }^{l \dot{{\alpha_{4}}}{\alpha_{3}}}$.\quad
Proof:\quad
From definition\\
$\Lambda ^{{\alpha_{1}} \dot{{\alpha_{2}}}{\alpha_{3}} \dot{{\alpha_{4}}}}=C^{{\alpha_{1}} {\beta_{1}}}~C^{\dot{{\alpha_{2}}} \dot{{\beta_{2}}}}~C^{{\alpha_{3}} {\beta_{3}}}~C^{\dot{{\alpha_{4}}} \dot{{\beta_{4}}}}~\Lambda _{{\beta_{1}} \dot{{\beta_{2}}}{\beta_{3}} \dot{{\beta_{4}}}}$,\quad
By using:\quad
$C^{\alpha \beta }=i ~\epsilon ^{\alpha \beta }$,\quad
$\Lambda _{{\beta_{1}} \dot{{\beta_{2}}}{\beta_{3}} \dot{{\beta_{4}}}}=\Lambda _{{kl}}~\sigma ^k{}_{{\beta_{1}} \dot{{\beta_{2}}}}~\sigma ^l{}_{{\beta_{3}} \dot{{\beta_{4}}}}$.

\item
$\Lambda _{\alpha  \dot{\beta }}{}^{\gamma }=-\Lambda ^{k \gamma }~\sigma _{k \alpha  \dot{\beta }}$.

\item
$\Lambda ^{\alpha  \dot{\beta }\gamma }~\partial _{\alpha  \dot{\beta }}=2~ {p\Lambda }^{\gamma }=2~\Lambda ^{k \gamma }~\partial _k$.\quad
Proof:\\
$\Lambda ^{\alpha  \dot{\beta }\gamma }~\partial _{\alpha  \dot{\beta }}=\Lambda ^{k \gamma }~\overline{\sigma }_k{}^{\dot{\beta }\alpha }~(-1)~\sigma ^l{}_{\alpha  \dot{\beta }}~\partial _l=(-1)(-2)~\Lambda ^{k \gamma }~\delta _k{}^l~\partial _l=2~\Lambda ^{k \gamma }~\partial _k=2~{p\Lambda }^{\gamma }$.

\item
$C^{{\alpha_{1}} {\alpha_{2}}}~\partial _{{\alpha_{1}} \dot{{\beta_{1}}}}~\partial _{{\alpha_{2}} \dot{{\beta_{2}}}}=\frac{1}{2}~C_{\dot{{\beta_{1}}}\dot{{\beta_{2}}}}~\partial ^{\gamma  \dot{\rho }}~\partial _{\gamma  \dot{\rho }}=C_{\dot{{\beta_{1}}}\dot{{\beta_{2}}}}~\square$.

\item
$\Lambda _{\alpha  \dot{\beta }\gamma }=i~ \Lambda ^k{}_{\gamma }~\sigma _{k \alpha  \dot{\beta }}$.

\item
$C^{{\alpha_{1}} {\alpha_{2}}}~\Lambda _{{\alpha_{1}} \dot{{\beta_{1}}}{\alpha_{2}} \dot{{\beta_{2}}}}=\Lambda ^2~C_{\dot{{\beta_{1}}}\dot{{\beta_{2}}}},\quad \Lambda ^2=\Lambda ^{{kl}}~\eta _{{kl}}$.\quad
Proof:\quad
$\Lambda _{{\alpha_{1}} \dot{{\beta_{1}}}{\alpha_{2}} \dot{{\beta_{2}}}}=\Lambda _{{kl}}~\sigma ^k{}_{{\alpha_{1}} \dot{{\beta_{1}}}}~\sigma ^l{}_{{\alpha_{2}} \dot{{\beta_{2}}}}$,\quad \\
$\epsilon ^{{\alpha_{1}} {\alpha_{2}}}~\Lambda _{{kl}}~\sigma ^k{}_{{\alpha_{1}} \dot{{\beta_{1}}}}~\sigma ^l{}_{{\alpha_{2}} \dot{{\beta_{2}}}}=\epsilon _{\dot{{\beta_{1}}}\dot{{\beta_{2}}}}~\Lambda ^2$,\quad
$C^{{\alpha_{1}} {\alpha_{2}}}=i~ \epsilon ^{{\alpha_{1}} {\alpha_{2}}}$,\quad 
$C_{\dot{{\beta_{1}}}\dot{{\beta_{2}}}}=i~ \epsilon _{\dot{{\beta_{1}}}\dot{{\beta_{2}}}}$.

\end{itemize}

\section{$D$ algebraic relations}

Definition:\quad
$\partial _{\alpha }=D_{\alpha }-i ~\overline{\theta }^{\dot{\alpha }}~\partial _{\alpha  \dot{\alpha }}$.

We have the following identities:
\begin{itemize}
\label{}
\item
$\partial _{\alpha }=D_{\alpha }+i~ \sigma ^k{}_{\alpha  \dot{\beta }}~\overline{\theta }^{\dot{\beta }}~\partial _k$.

\item
$\int d^4x~ d^4\theta  ~d^{{c_{1}} {c_{2}} {c_{3}}}~\overline{\Gamma }^{\dot{{\alpha_{1}}}{c_{1}}}~\overline{W}^{\dot{{\alpha_{2}}}{c_{2}}}~\overline{W}^{\dot{{\alpha_{3}}}{c_{3}}}=0$.

\item
$D^2\overline{D}_{\dot{\alpha }}\overline{W}_{\dot{\beta }}^{c}=0$,\quad
$D_{\alpha }\overline{D}_{\dot{\beta }}\overline{W}_{\dot{\gamma }}^{c}=i~\partial _{\alpha  \dot{\beta }}\overline{W}_{\dot{\gamma }}^{c}$,\quad
We have used:\quad
$\left\{D_{\alpha },\overline{D}_{\dot{\beta }}\right\}=i ~\partial _{\alpha  \dot{\beta }}$.

\end{itemize}

\section{Supergauge transformation}

Definition of supergauge transformation :\quad
$\overline{\nabla }_A\to  e^{i \overline{\Lambda }}_* *~\overline{\nabla }_A*~e^{-i \overline{\Lambda }}_*$,\quad 
where $D_{\alpha }\overline{\Lambda }=0$.

We have the infinitesimal supergauge transformation:\quad
$\delta~  \overline{\Gamma }_{\dot{\alpha }}=\overline{D}_{\dot{\alpha }}\overline{\Lambda }+i~\left[~\overline{\Lambda },~\overline{\Gamma }_{\dot{\alpha }}~\right]_*$,\quad
$\delta~  \overline{\Gamma }_{\alpha  \dot{\beta }}=\partial _{\alpha  \dot{\beta }}\overline{\Lambda }+i~\left[~\overline{\Lambda },~\overline{\Gamma }_{\alpha  \dot{\beta }}~\right]_*$,\quad
$\delta~  \overline{W}_{\dot{\alpha }}=i~\left[~\overline{\Lambda },~\overline{W}_{\dot{\alpha }}~\right]_*$.\quad
We have used:\quad
$\overline{\nabla }_{\dot{\alpha }}=\overline{D}_{\dot{\alpha }}-i~\overline{\Gamma }_{\dot{\alpha }}$, \quad 
$e^X~Y~ e^{-X}=Y+[X,~Y]+O^2(X)$.

\section{Noncommutative spacetime}

\begin{itemize}
\label{}
\item
Definition: \quad
$\left[x^k,~\theta ^{\alpha }\right]=i~ \Lambda ^{{k\alpha }}$.

\item
We have the relation:\quad
$\left[x^{\alpha  \dot{\beta }},~\theta ^{\gamma }\right]=\frac{i}{2}~\Lambda ^{\alpha  \dot{\beta }\gamma }$.
Proof:\quad 
By using
$\left[x^k,~\theta ^{\gamma }\right]=i~ \Lambda ^{{k\alpha }}$,\quad
we have
$\frac{1}{2}~\sigma _k{}^{\alpha  \dot{\beta }}~\left[x^k,~\theta ^{\gamma }\right]=\frac{i}{2}~\sigma _k{}^{\alpha  \dot{\beta }}~ \Lambda ^{{k\alpha }}$,\quad
then we have
$\left[x^{\alpha  \dot{\beta }},~\theta ^{\gamma }\right]=\frac{i}{2}~\Lambda ^{\alpha  \dot{\beta }\gamma }$.

\end{itemize}

\section{BFNC star product}

The star product corresponding to
$\left[x^k,~\theta ^{\alpha }\right]=i~ \Lambda ^{{k\alpha }}$
is:\\
$F*G=\mu \left\{\exp \left[\frac{i}{2}~\Lambda ^{{k\alpha }}\left(\partial _k\otimes \partial _{\alpha }-\partial _{\alpha }\otimes \partial _k\right)\right]\triangleright (F\otimes G)\right\}$.\quad
At first order of $\Lambda ^{{k\alpha }}$:
\begin{eqnarray}
\label{}
F*G&=&F ~G+(-1)^{|F|}~\frac{i}{2}~\Lambda ^{{k\alpha }}~\left(\partial _kF\right)~\left(\partial _{\alpha }G\right)-\frac{i}{2}~\Lambda ^{{k\alpha }}~\left(\partial _{\alpha }F\right)~\left(\partial _kG\right)\nonumber\\
&=&F ~G+(-1)^{|F|}~\frac{i}{4}~\Lambda ^{\beta \dot{\gamma }\alpha }~\left(\partial _{\beta \dot{\gamma }}F\right)~\left(\partial _{\alpha }G\right)-\frac{i}{4}~\Lambda ^{\beta \dot{\gamma }\alpha }~\left(\partial _{\alpha }F\right)~\left(\partial _{\beta \dot{\gamma }}G\right)\nonumber\\
&=&F ~G+(-1)^{|F|}~\frac{i}{2}~\left({p\Lambda }^{\alpha }F\right)~\left(\partial _{\alpha }G\right)+(-1)^{|F|}~\frac{i}{2}~\left(\partial _{\alpha }F\right)~\left({p\Lambda }^{\alpha }G\right)
\end{eqnarray}

By using BFNC star product we have,
\begin{itemize}
\label{}
\item
$x^k*\theta ^{\alpha }-\theta ^{\alpha }*x^k=i ~\Lambda ^{k \alpha }$.\quad
Proof: \quad
By using 
${p\Lambda }^{\alpha }~x^k=\Lambda ^{{k\alpha }}$
we have
$x^k*\theta ^{\alpha }=x^k~\theta ^{\alpha }+\frac{i}{2}~\Lambda ^{{k\alpha }}$ 
and
$\theta ^{\alpha }*x^k=\theta ^{\alpha }~x^k-\frac{i}{2}~\Lambda ^{{k\alpha }}$,\quad
then we have 
$x^k*\theta ^{\alpha }-\theta ^{\alpha }*x^k=i ~\Lambda ^{k \alpha }$.

\item
$e^{-i~ \Pi _1~Z}*e^{-i~ \Pi _2~Z}=e^{\Pi _1~\land ~\Pi _2}~e^{-i ~\Pi _1~Z}~e^{-i ~\Pi _2~Z}$,\\
where \quad
$\Pi _1\land \Pi _2=-\frac{i}{2}~\left(\Lambda ^{{k\alpha }}~p_{1k}~\pi _{2\alpha }-\Lambda ^{{k\alpha }}~p_{2k}~\pi _{1\alpha }\right)$.\quad
Proof: \quad
By using
$\partial _k\left(-i ~\Pi _1~ Z\right)=-i ~p_{1k}$
and
$\partial _{\alpha }\left(-i~ \Pi _1 ~Z\right)=-i~ \pi _{1\alpha }$.

\item
$[M,N]_*=\frac{1}{2}~d^{{ABC}}~\left[~M^A,~N^B~\right]_*~T^C+\frac{i}{2}~f^{{ABC}}~\left\{~M^A,~N^B~\right\}_*~T^C$,\quad
where $M=M^A~T^A$,\quad $N=N^A~T^A$,\quad $\left[T^A,~T^B\right]=i ~f^{{ABC}}~T^C,\quad \left\{T^A,~T^B\right\}=d^{{ABC}}~T^C$.

\end{itemize}

\section{Divergent structure}

\begin{itemize}
\label{}
\item
At the first order of $\Lambda ^{{k\alpha }}$, the following  Feynman diagrams are not divergent:  $S_1~S_2{}^n$ $(n\geq  3)$.\quad
Proof:\quad
$S_1~S_2~S_2~S_2\to  d^4p ~{\Lambda p}~ \frac{1}{\left(p^2+m \overline{m}\right)^4}~\left(\widetilde{D}^2\widetilde{\overline{D}} \right)~\widetilde{\overline{D}} ~p ~p \sim  \frac{p^7}{p^8}$.

\item
The coefficient at order $n$ is $\frac{1}{n}~2^{n-1}$.
\end{itemize}

\section{The derivative operators}

Definition:\quad 
$\overline{\eth}_{\dot{\alpha }}~\overline{\pi }^{\dot{\beta }}=\delta _{\dot{\alpha }}{}^{\dot{\beta }}$.

We have the following identities:
\begin{itemize}
\label{}
\item
$\overline{\eth}^{\dot{\alpha }}~\overline{\pi }_{\dot{\beta }}=-\delta ^{\dot{\alpha }}{}_{\dot{\beta }}$.\quad
Proof:\quad
$\overline{\eth}^{\dot{\alpha }}~\overline{\pi }_{\dot{\beta }}=C^{\dot{\alpha }\dot{\gamma }}~C_{\dot{\rho }\dot{\beta }}~\overline{\eth}_{\dot{\gamma }}~\overline{\pi }^{\dot{\rho }}=C^{\dot{\alpha }\dot{\gamma }}~C_{\dot{\rho }\dot{\beta }}~\delta _{\dot{\gamma }}{}^{\dot{\rho }}=-\delta ^{\dot{\alpha }}{}_{\dot{\beta }}$.

\item
$\overline{\eth}^{\dot{\alpha }}\left(i~ \overline{\pi }^{\dot{\beta }}~\overline{\theta }_{\dot{\beta }}\right)=i~ \overline{\theta }^{\dot{\alpha }}$.\quad
Proof:\quad
$\overline{\eth}^{\dot{\alpha }}\left(i ~\overline{\pi }^{\dot{\beta }}~\overline{\theta }_{\dot{\beta }}\right)=(-i)~\overline{\eth}^{\dot{\alpha }}\left(\overline{\pi }_{\dot{\beta }}~\overline{\theta }^{\dot{\beta }}\right)=(-i)(-1)~\delta ^{\dot{\alpha }}{}_{\dot{\beta }}~\overline{\theta }^{\dot{\beta }}=i ~\overline{\theta }^{\dot{\alpha }}$.

\item
${p\Lambda }_{\alpha }\left(X^{\beta }~X_{\beta }\right)=-2~X^{\beta }\left({p\Lambda }_{\alpha }~X_{\beta }\right)=2\left({p\Lambda }_{\alpha }~X^{\beta }\right)X_{\beta }$.\quad
Proof:\quad
${p\Lambda }_{\alpha }\left(X^{\beta }~X_{\beta }\right)=\left({p\Lambda }_{\alpha }~X^{\beta }\right)X_{\beta }-X^{\beta }\left({p\Lambda }_{\alpha }~X_{\beta }\right)=X_{\beta }\left({p\Lambda }_{\alpha }~X^{\beta }\right)-X^{\beta }\left({p\Lambda }_{\alpha }~X_{\beta }\right)=-2~X^{\beta }\left({p\Lambda }_{\alpha }~X_{\beta }\right)=2~\left({p\Lambda }_{\alpha }~X^{\beta }\right)X_{\beta }$.

\item
$\partial _k\left(X^{\beta }~X_{\beta }\right)=2~ X^{\beta }\left(\partial _k~X_{\beta }\right)$.\quad
Proof:\quad
$\partial _k\left(X^{\beta }~X_{\beta }\right)=\left(\partial _k~X^{\beta }\right)X_{\beta }+X^{\beta }\left(\partial _k~X_{\beta }\right)=2~\left(\partial _k~X^{\beta }\right)X_{\beta }$.

\end{itemize}

\section{Fierz identities}

Definition:\quad
$\overline{W}^{{cc}}=\frac{1}{2}~\overline{W}^{\dot{\gamma }c}~\overline{W}_{\dot{\gamma }}^c$,\quad
$X^2=\frac{1}{2}~X^{\alpha }~X_{\alpha }$,\quad
$X^{\alpha }=C^{\alpha \beta }~X_{\beta }$,\quad 
$C_{\alpha \beta }=-C_{\beta \alpha }$,\quad 
$C^{\alpha \beta }=-C^{\beta \alpha }$,\quad
$p^2=\frac{1}{2}~p^{\alpha  \dot{\beta }}~p_{\alpha  \dot{\beta }}$,\quad
$D^2~\overline{\Gamma }^{\dot{\alpha }A}=\overline{W}^{\dot{\alpha }A}$,\quad 
$\overline{\Gamma }^{\alpha  \dot{\beta }A}=-i~ D^{\alpha }~\overline{\Gamma }^{\dot{\beta }A}$.

We have the following identities:
\begin{itemize}
\label{}
\item
$\pi _{\alpha }~\pi _{\beta }~\pi _{\gamma }=0$,\quad 
${\Lambda p}^{\alpha }~{\Lambda p}^{\beta }~{\Lambda p}^{\gamma }=0$.

\item
$\overline{W}_{\dot{\alpha }}^c~\overline{W}_{\dot{\beta }}^c=\frac{1}{2}~C_{\dot{\beta }\dot{\alpha }}~\overline{W}^{\dot{\gamma }c}~\overline{W}_{\dot{\gamma }}^c=C_{\dot{\beta }\dot{\alpha }}~\overline{W}^{{cc}}$.

\item
$X_{\alpha }~X_{\beta }=C_{\beta \alpha }~X^2$,\quad 
$X^{\alpha }~X^{\beta }=C^{\beta \alpha }~X^2,\quad 
X_{\alpha }~X^{\beta }=-\delta _{\alpha }{}^{\beta }~X^2$,\quad 
$X^{\alpha }~X_{\beta }=\delta ^{\alpha }{}_{\beta }~X^2$.

\item
$X_{\alpha }=X^{\beta }~C_{\beta \alpha }$,\quad $C^{\alpha \beta }~C_{\beta \gamma }=-\delta ^{\alpha }{}_{\gamma }$,\quad $\delta _{\alpha }{}^{\alpha }=2$,\quad $\delta ^{\alpha }{}_{\alpha }=2$.

\item
$C^{\alpha \beta }~p_{\alpha  \dot{\gamma }}~p_{\beta  \dot{\rho }}=C_{\dot{\gamma }\dot{\rho }}~p^2$.

\item
$X_{\alpha }~Y^{\alpha }=-X^{\alpha }~Y_{\alpha }$.

\end{itemize}

\end{CJK}

\newpage

\begin{thebibliography}{999}


\bibitem{Douglas:2001ba} 
  M.~R.~Douglas and N.~A.~Nekrasov,
  ``Noncommutative field theory,''
  Rev.\ Mod.\ Phys.\  {\bf 73}, 977 (2001)
  [arXiv:hep-th/0106048].
  
\bibitem{Szabo:2001kg} 
  R.~J.~Szabo,
  ``Quantum field theory on noncommutative spaces,''
  Phys.\ Rept.\  {\bf 378}, 207 (2003)
  [arXiv:hep-th/0109162].





\bibitem{Seiberg:1999vs} 
  N.~Seiberg and E.~Witten,
  ``String theory and noncommutative geometry,''
  JHEP {\bf 9909}, 032 (1999)
  [arXiv:hep-th/9908142].

\bibitem{de Boer:2003dn} 
  J.~de Boer, P.~A.~Grassi, and P.~van Nieuwenhuizen,
  ``Noncommutative superspace from string theory,''
  Phys.\ Lett.\ B {\bf 574}, 98 (2003)
  [arXiv:hep-th/0302078].

\bibitem{Ooguri:2003qp} 
  H.~Ooguri and C.~Vafa,
  ``The C-deformation of gluino and nonplanar diagrams,''
  Adv.\ Theor.\ Math.\ Phys.\  {\bf 7}, 53 (2003)
  [arXiv:hep-th/0302109].

\bibitem{Ooguri:2003tt} 
  H.~Ooguri and C.~Vafa,
  ``Gravity induced C-deformation,''
  Adv.\ Theor.\ Math.\ Phys.\  {\bf 7}, 405 (2004)
  [arXiv:hep-th/0303063].

\bibitem{Berkovits:2003kj} 
  N.~Berkovits and N.~Seiberg,
  ``Superstrings in graviphoton background and N=1/2 + 3/2 supersymmetry,''
  JHEP {\bf 0307}, 010 (2003)
  [arXiv:hep-th/0306226].



\bibitem{Seiberg:2003yz} 
  N.~Seiberg,
 ``Noncommutative superspace, N = 1/2 supersymmetry, field theory and string theory,''
  JHEP {\bf 0306}, 010 (2003)
  [arXiv:hep-th/0305248].



\bibitem{Britto:2003aj} 
  R.~Britto, B.~Feng and S.~J.~Rey,
 ``Deformed superspace, N = 1/2 supersymmetry and nonrenormalization theorems,''
  JHEP {\bf 0307}, 067 (2003)
  [arXiv:hep-th/0306215].


\bibitem{Grisaru:2003fd} 
  M.~T.~Grisaru, S.~Penati, and A.~Romagnoni,
``Two loop renormalization for nonanticommutative N = 1/2 supersymmetric WZ model,''
  JHEP {\bf 0308}, 003 (2003)
  [arXiv:hep-th/0307099].


\bibitem{Britto:2003kg} 
  R.~Britto and B.~Feng,
 ``N=1/2 Wess-Zumino model is renormalizable,''
  Phys.\ Rev.\ Lett.\  {\bf 91}, 201601 (2003)
  [arXiv:hep-th/0307165].


\bibitem{Romagnoni:2003xt} 
  A.~Romagnoni,
 ``Renormalizability of N=1/2 Wess-Zumino model in superspace,''
  JHEP {\bf 0310}, 016 (2003)
  [arXiv:hep-th/0307209].



\bibitem{Grisaru:2005we} 
  M.~T.~Grisaru, S.~Penati and A.~Romagnoni,
  ``Non(anti)commutative sym theory: Renormalization in superspace,''
  JHEP {\bf 0602}, 043 (2006)
  [hep-th/0510175].



\bibitem{Miao:2013a} 
  Y.~G.~Miao and X.~D.~Wang,
  ``One-loop renormalizable Wess-Zumino model on bosonic-fermionic noncommutative superspace,''
  Phys. Rev. D {\bf 90},  045036 (2014) [arXiv:1403.4705 [hep-th]].


\bibitem{Miao:2014mia} 
  Y.~G.~Miao and X.~D.~Wang,
  ``All-Loop Renormalizable Wess-Zumino Model on Bosonic-Fermionic Noncommutative Superspace,''
  Phys.\ Rev.\ D {\bf 91}, no. 2, 025016 (2015)
  [arXiv:1403.5046 [hep-th]].



\bibitem{Gates:1983nr} 
  S.~J.~Gates, M.~T.~Grisaru, M.~Rocek, and W.~Siegel,
  ``Superspace or one thousand and one lessons in supersymmetry,''
  Front.\ Phys.\  {\bf 58}, 1 (1983)
  [arXiv:hep-th/0108200].


\bibitem{Penati:2004gh} 
  S.~Penati and A.~Romagnoni,
  ``Covariant quantization of N = 1/2 SYM theories and supergauge invariance,''
  JHEP {\bf 0502}, 064 (2005)
  [hep-th/0412041].




\end{thebibliography}
\end{document}